\documentclass[sigconf]{acmart}
\AtBeginDocument{%
  }

\setcopyright{acmlicensed}
\copyrightyear{2026}
\acmYear{2026}
\setcopyright{cc}
\setcctype{by}
\acmConference[CHI EA '26]{Extended Abstracts of the 2026 CHI Conference on Human Factors in Computing Systems}{April 13--17, 2026}{Barcelona, Spain}
\acmBooktitle{Extended Abstracts of the 2026 CHI Conference on Human Factors in Computing Systems (CHI EA '26), April 13--17, 2026, Barcelona, Spain}
\acmDOI{10.1145/3772363.3798690}
\acmISBN{979-8-4007-2281-3/2026/04}

\usepackage{multirow}
\usepackage{longtable}
\usepackage{appendix}

\begin{document}

\title{52-Hz Whale Song: An Embodied VR Experience for Exploring Misunderstanding and Empathy}

\author{Yibo Meng}
\affiliation{%
  \institution{Tsinghua University}
  \city{Beijing}
  %\state{Washington}
  \country{China}
}
\email{mengyb22@tsinghua.org.cn}

\author{Bingyi Liu}
\affiliation{%
  \institution{University of Michigan, Ann Arbor}
  \city{Ann Arbor}
  \state{Michigan}
  \country{United State}
}
\email{bingyi@umich.edu}

\author{Ruiqi Chen}
\affiliation{%
  \institution{University of Washington}
  \city{Seattle}
  \state{Washington}
  \country{United States}
}
\email{ruiqich@uw.edu}

\author{Xin Chen}
\affiliation{%
  \institution{Universidad Politécnica de Madrid}
  \city{Madrid}
  %\state{Washington}
  \country{Spain}
}
\email{xin.c@alumnos.upm.es}

\author{Yan Guan}
\affiliation{%
  \institution{Tsinghua University}
  \city{Beijing}
  %\state{Washington}
  \country{China}
}
\email{guany@tsinghua.edu.cn}
\renewcommand{\shortauthors}{XXX et al.}

\begin{abstract}
Experiences of being misunderstood often stem not from a lack of voice, but from mismatches between how individuals express themselves and how others listen. Such communicative mismatches arise across many social settings, including situations involving linguistic and cultural displacement. While prior HCI research has explored empathy through virtual reality, many approaches rely on narrative explanation, positioning users as observers rather than embodied participants. We present 52-Hz Whale Song, an embodied VR experience that explores miscommunication through metaphor and perspective-shifting. Inspired by the real-world ``52-Hz whale,'' whose calls are not responded to by others, the experience uses this phenomenon as an experiential lens on communicative mismatch rather than representing any specific social group. Players progress through a three-act arc that moves from failed communication to agency and ultimately to mediation. A preliminary mixed-methods study ($N = 30$) suggests increased perspective-taking and reduced self-reported social distance in immigrant-related situations. This work highlights how embodied metaphor and role-shifting can support empathic engagement and offers transferable design insights for empathy-oriented interactive systems.

\end{abstract}

\begin{CCSXML}
<ccs2012>
   <concept>
       <concept_id>10003120.10003121.10003128</concept_id>
       <concept_desc>Human-centered computing~Interaction techniques</concept_desc>
       <concept_significance>500</concept_significance>
       </concept>
   <concept>
       <concept_id>10003120.10003121.10003125</concept_id>
       <concept_desc>Human-centered computing~Interaction devices</concept_desc>
       <concept_significance>500</concept_significance>
       </concept>
   <concept>
       <concept_id>10010405.10010469.10010474</concept_id>
       <concept_desc>Applied computing~Media arts</concept_desc>
       <concept_significance>500</concept_significance>
       </concept>
   <concept>
       <concept_id>10003456.10010927</concept_id>
       <concept_desc>Social and professional topics~User characteristics</concept_desc>
       <concept_significance>500</concept_significance>
       </concept>
 </ccs2012>
\end{CCSXML}

\ccsdesc[500]{Human-centered computing~Interaction techniques}
\ccsdesc[500]{Human-centered computing~Interaction devices}
\ccsdesc[500]{Applied computing~Media arts}
\ccsdesc[500]{Social and professional topics~User characteristics}

\keywords{Empathy Enhancement, Virtual Reality, Multimodal Interaction, Embodied Interaction, Social Inclusion, Experiential Learning}

\begin{teaserfigure}
  \includegraphics[width=\textwidth]{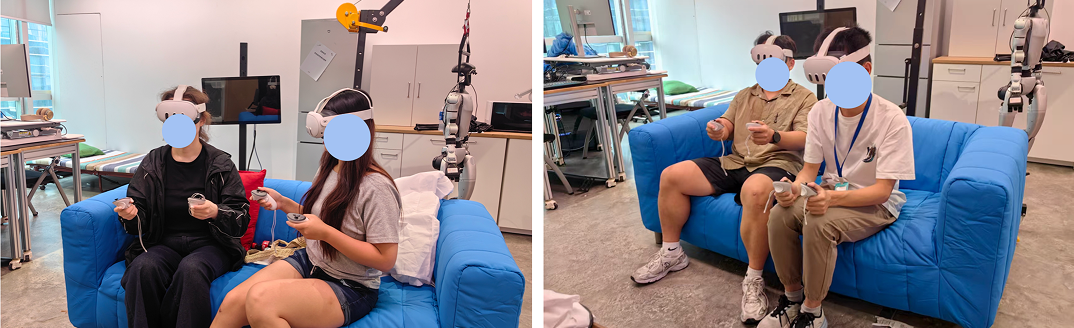}
 \caption{Participants experiencing \textit{52-Hz Whale Song} during the study session (photo used with permission).}
  \label{fig0}
\end{teaserfigure}

\maketitle

\section{Introduction}

Electronic games and virtual reality (VR) have increasingly been recognized within HCI not only as entertainment media, but as powerful interactive systems capable of shaping emotional experience, social cognition, and learning \cite{wei2025systematic,bujic2025virtually}. Prior work demonstrates their potential to support mental wellbeing, embodied learning, and emotional regulation \cite{bujic2025virtually, chen2023design,liang2025user,ozkaya2025investigating,kirchner2024outplay}, as well as to foster cross-cultural understanding and social connection through immersive interaction \cite{fernandez2025breaking,zhao2025immersive,wagener2025togetherreflect}. In particular, VR’s affordances for immersion, presence, and embodiment have positioned it as a promising medium for empathy-oriented design.

Within this context, empathy—encompassing affective resonance and cognitive perspective-taking—has become a central concern in games and VR research \cite{ju2025haptic,gencc2024situating, meng2026mistyforest}. In this work, we focus on empathy as short-term shifts in perspective-taking and self-reported social distance, with mediation framed as action rather than passive sympathy. Existing systems explore immersive storytelling, social VR, and multimodal interaction to deepen interpersonal understanding, including multimodal feedback for perspective-taking, co-presence in shared virtual environments, and inclusive designs for marginalized communities \cite{ju2025haptic,shrestha2025virtual,wang2025facilitating,desnoyers2025being,xie2025vrcaptions,kong2025working,choi2025aacesstalk,bei2024starrescue}. Many of these efforts align with the “Double Empathy” framework, which reframes communication barriers as bidirectional misunderstandings rather than deficits located in a single group \cite{kong2025working}.

However, despite these advances, important gaps remain. First, many empathy-oriented systems focus on specific populations — particularly neurodivergent users—while the experience of being misunderstood is not exclusive to any one group. Rather, it is a broadly shared interactional condition that can emerge in contexts of linguistic, cultural, or social unfamiliarity, such as navigating a new country, entering an unfamiliar community, or operating across differing communicative norms. Second, while numerous interventions provide functional assistance or explanatory narratives, fewer designs support deeper experiential transformation through embodied interaction and artistic metaphor \cite{choi2025aacesstalk,park2025lessons}. As a result, users often learn \emph{about} misunderstanding rather than \emph{feeling} its emotional and bodily consequences. Methodologically, experience-centered designs remain underexplored. In particular, few systems integrate visual, auditory, and haptic modalities to render the transition from communication failure to active mediation as a cohesive embodied experience \cite{smith2025archaeological,desnoyers2025being}.

In response, we present \textit{52-Hz Whale Song}, a VR-based multimodal interactive experience that uses embodied metaphor and perspective-shifting to make the experience of being misunderstood tangible. Inspired by the scientific phenomenon of the “52-Hz whale”—a solitary whale whose calls occur at a frequency other whales do not respond to—the experience reframes miscommunication as an interactional and bodily problem rather than a narrative explanation. Importantly, the metaphor is not intended to represent any social group as a monolith, but to offer an experiential lens on communicative mismatch that participants may map onto diverse real-world situations. The experience unfolds across three acts: embodying isolation as the whale, confronting crisis alone, and ultimately becoming a pink dolphin who mediates understanding between others. Through this arc, players do not simply observe exclusion, but encounter it through their own actions, sensory feedback, and shifting agency. Rather than waiting to be understood, they learn—through embodied interaction and procedural rhetoric—that understanding requires active effort and translation.

This work contributes (1) \textit{52-Hz Whale Song}, an experience-first VR design that makes being misunderstood felt through embodied, multimodal interaction; (2) a role-shifting interaction pattern that reframes empathy as active mediation; and (3) initial evidence of increased perspective-taking and reduced self-reported social distance in immigrant-related contexts.

Together, these contributions point to role-shifting as a viable design strategy for empathy-oriented VR systems, enabling users to move from experiencing communicative breakdown to actively mediating understanding.

\section{The Experience}

\begin{figure*}[t]
    \centering
    \includegraphics[width=\textwidth]{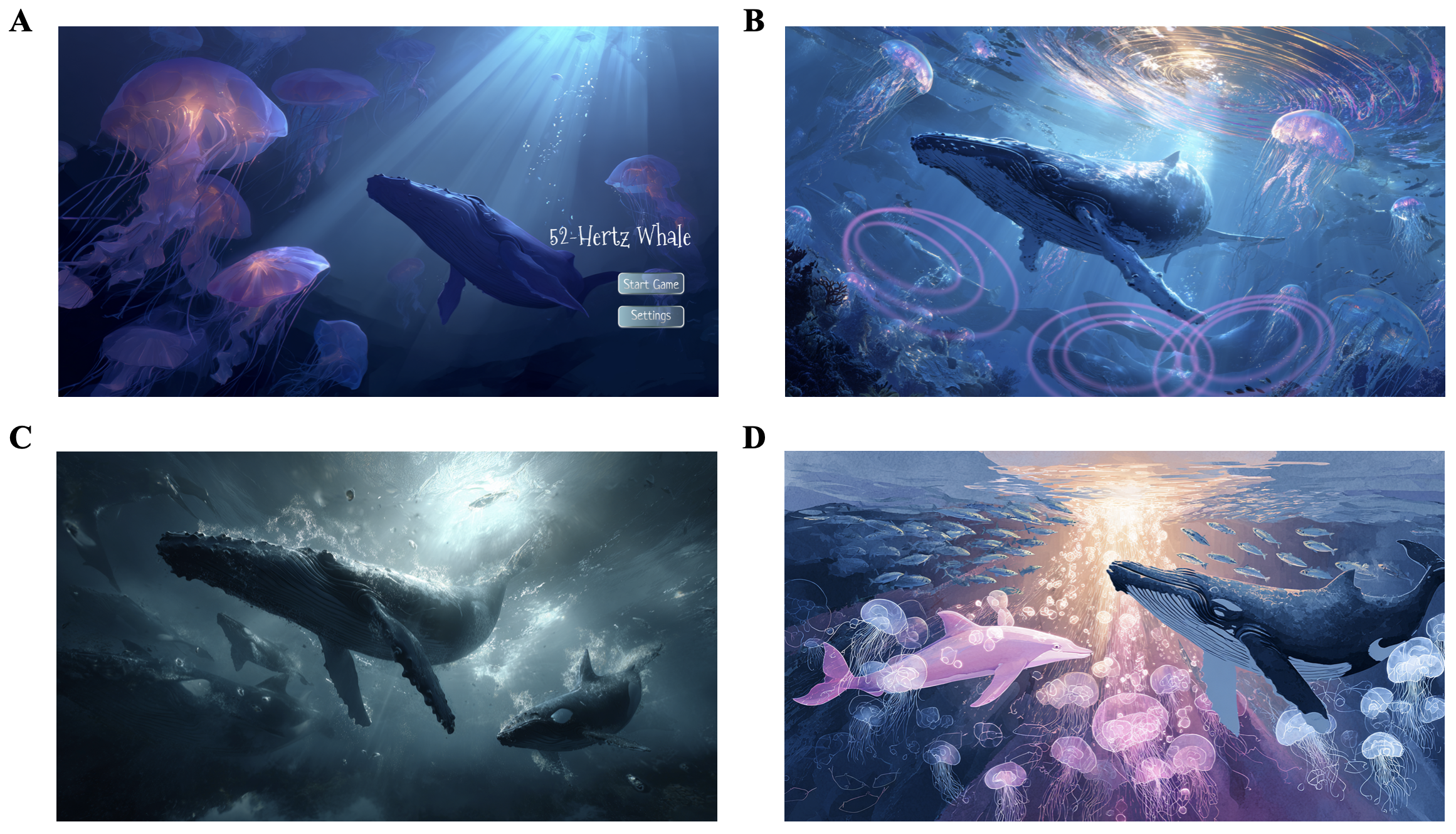}
    \caption{Overview of the experience arc and interactional roles in \textit{52-Hz Whale Song}. 
    (A) Entry interface establishing emotional tone; 
    (B) communication attempts visualized as sonic waves and rejected by an invisible barrier; 
    (C) survival crisis reinforcing embodied isolation; 
    (D) role shift to a mediating agent translating experiences and bridging understanding.}
    \label{fig:experience-arc}
\end{figure*}

\textit{52-Hz Whale Song} is an experience-first VR narrative structured as a three-act arc that moves players from isolation to agency and ultimately to mediation. Rather than explaining miscommunication through narrative exposition, the experience renders it tangible through embodied interaction, sensory feedback, and role-shifting.

\textbf{Act I: Being Misunderstood.}
Players embody a whale whose vocalizations appear as golden sound waves. Attempts to communicate with a pod of whales consistently fail as these signals collide with an invisible sonic barrier and dissipate. Players can move, gesture, and vocalize, yet remain unheard. This act establishes miscommunication as a bodily and spatial experience characterized by frustration and exclusion.

\textbf{Act II: Confronting Isolation.}
A crisis forces players to face danger alone. Calling for help remains ineffective, and escape is impossible. Progress requires players to act through their own embodied effort, shifting the experience from passive exclusion to self-assertion without resolving the communication barrier. Emotionally, the experience transitions from helplessness to resilience.

\textbf{Act III: Becoming the Bridge.}
Players assume the role of a dolphin capable of perceiving and translating the whale’s signals. By listening to and re-expressing fragments of prior experiences, players enable communication to succeed through mediation. The act culminates in reconciliation, emphasizing that understanding requires effort and responsibility.

Across all acts, interaction relies on full-body movement, voice input, spatial audio, and haptic feedback. By shifting players from the position of the misunderstood to that of the mediator, the experience reframes empathy from passive perspective-taking to active bridge-building.

\section{Design Approach}

\subsection{Design Rationale}

The design of \textit{52-Hz Whale Song} treats misunderstanding as an interactional and embodied condition rather than a narrative theme. Instead of explaining social exclusion through exposition, the experience allows players to encounter communication failure through repeated action, sensory feedback, and resistance. Grounded in experiential learning, meaning emerges through doing and reflecting rather than receiving information \cite{kolb2014experiential}.

We prioritize embodied failure over representational explanation. Early communication attempts are intentionally designed to fail: players’ vocalizations and gestures generate visible and haptic responses that are nonetheless rejected. By mapping miscommunication to physical obstruction and sensory resistance, misunderstanding is shifted from an abstract social concept to a felt bodily experience.

Metaphor serves as a generative and protective design strategy. Marine life and acoustic frequency provide emotional distance from real-world identities while remaining transferable to lived social experiences. This framing enables players to relate the experience to immigrants, neurodivergent individuals, or other marginalized groups without explicit instruction, supporting reflection without didactic framing.

The experience further centers on role-shifting—from the misunderstood subject to the mediating agent. Rather than leaving players in passive empathy or pity, the design culminates in a translator role that reframes empathy as responsibility and action, aligning affective resonance with cognitive perspective-taking.

Finally, we employ procedural rhetoric to embed values directly into interaction rules \cite{bogost2010persuasive}. The system’s mechanics perform the argument that being unheard reflects a mismatch of communicative conditions rather than a lack of voice, and that understanding requires active effort.

\subsection{System and Interaction Design}

To operationalize this experience-first approach, the system implements three core interaction mechanisms that work together across the three acts.

\textbf{Sonic Visualization (Voice-to-Wave Mapping).}
Player vocalizations are transformed into dynamic visual waveforms with synchronized haptic feedback, making communication attempts visible and tangible even when they are not acknowledged.

\textbf{Embodied Communication Barrier (Sonic Net).}
Miscommunication is materialized through a volumetric blocking field that repels waves and constrains proximity. Continuous vibration, attenuation, and spatial distortion create a persistent yet reversible sense of obstruction, mirroring subtle social exclusion.

\textbf{Memory-Based Translation and Bridge-Building.}
Emotionally salient memory fragments captured earlier are reused in the final act to translate the whale’s experience into an intelligible form through hybrid sonic waves. A complementary buffer mechanic requires players to physically position themselves between groups to block exclusionary signals, transforming mediation into embodied labor.

The system is implemented in Unity and deployed on standalone VR headsets, using a modular architecture that separates narrative control, interaction logic, and multimodal feedback rendering.

\section{Preliminary Evaluation}
To explore the empathic potential of this experience-first design, we conducted a preliminary evaluation combining quantitative measures and qualitative reflections.

\subsection{Experimental Setup}

To examine whether the experience could transfer beyond its marine metaphor, we conducted a between-subjects study with 30 adult participants randomly assigned to a VR experience or control condition. Immigration-related scenarios were used as a socially salient test case of communicative mismatch rather than as a primary target population. This framing allowed us to explore whether embodied metaphor and role-shifting might generalize to real-world contexts involving social distance. The study followed a baseline–intervention–post-study procedure summarized in Table~\ref{tab:procedure}.

\begin{table}[t]
\centering
\caption{Overview of the experimental procedure.}
\begin{tabular}{p{0.2\linewidth} p{0.7\linewidth}}
\hline
\textbf{Stage} & \textbf{Procedure} \\
\hline
T0 (Baseline) 
& All participants viewed a short documentary on new immigrant experiences, completed baseline empathy and social distance measures, and participated in brief interviews. \\

T1 (Intervention) 
& \textbf{Experimental group:} Full VR playthrough of \textit{52-Hz Whale Song}. \newline
  \textbf{Control group:} No VR intervention. \\

T2 (Post-study) 
& \textbf{All participants:} Post-study measures and semi-structured interviews. \newline
  \textbf{Experimental group only:} Usability feedback (SUS). \\
\hline
\end{tabular}
\label{tab:procedure}
\end{table}

\subsection{Findings}

\textbf{Quantitative Findings.} As an exploratory, single-session evaluation with a modest sample size, the following results should be interpreted as preliminary and suggestive rather than conclusive.

\begin{figure}[t]
    \centering
    \includegraphics[width=\linewidth]{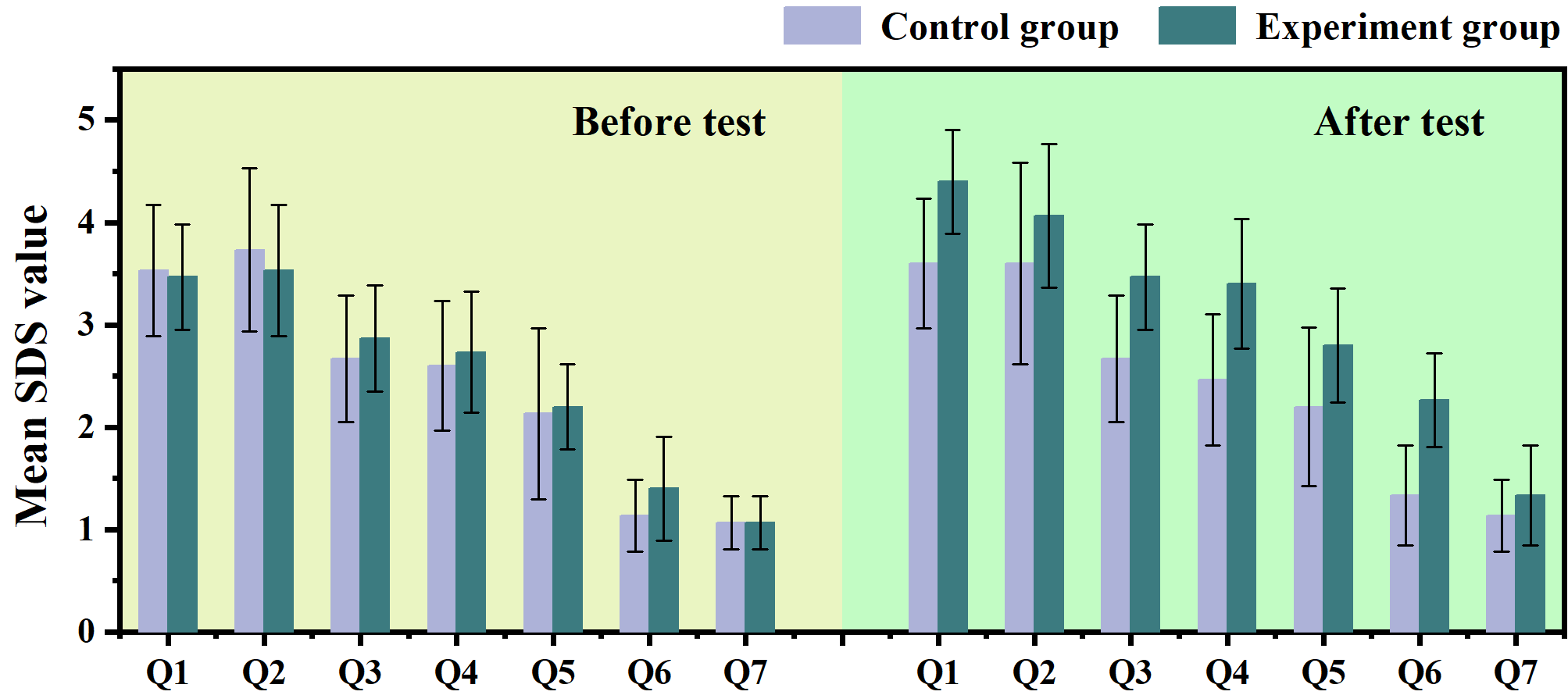}
    \caption{Changes in Social Distance Scale (SDS) scores in immigrant-related contexts for experimental and control groups (mean values).}

    \label{fig:sds}
\end{figure}

Participants who experienced \textit{52-Hz Whale Song} reported larger post-intervention increases in social acceptance in immigrant-related contexts than those in the control condition. In particular, post-intervention Social Distance Scale (SDS) scores showed significant between-group differences on several dimensions related to public and interpersonal relationships (e.g., Q1, Q4, Q6; all $p < .001$, $d = 1.42$--$1.94$), while the control group exhibited minimal change (Fig.~\ref{fig:sds}). Empathy-related measures showed a similar pattern. Relative to baseline, participants in the experimental group demonstrated significant gains in IRI Perspective-Taking and Fantasy subscales (both $p < .001$), with other subscales showing consistent directional improvements. Together, these preliminary findings suggest that the experience may support short-term perspective-taking and reductions in self-reported social distance beyond informational exposure alone.

\textbf{Qualitative interviews} help explain how these changes emerged. Participants described miscommunication as a bodily and emotional experience rather than an abstract concept, often referencing the repeated failure of communication attempts. The role transition from the isolated whale to the mediating dolphin was frequently cited as a turning point, reframing empathy from passive sympathy to active responsibility. Although the experience relied on marine-life metaphor, participants consistently mapped the narrative onto real-world contexts such as immigration and social marginalization, indicating transfer beyond the virtual setting. We emphasize that the 52-Hz whale functions as an experiential metaphor for communicative mismatch rather than a representation of any specific social group.
Together, these findings suggest that empathy-oriented VR systems can benefit from role-shifting designs that help users experience miscommunication as something shaped by interaction, rather than as a personal flaw.

\section{Discussion and Implications}

This work explores how experience-first, embodied interaction may support short-term empathic engagement in contexts involving social misunderstanding. Rather than conveying empathy through explanation or narrative exposition, \textit{52-Hz Whale Song} operationalizes misunderstanding as an interactional condition that players must bodily encounter and later mediate. The observed short-term shifts in social acceptance and perspective-taking suggest that empathic engagement, as operationalized in this study, may arise from inhabiting the constraints under which communication fails. Empathy may therefore emerge not only from understanding others’ stories, but from experiencing interactional breakdown directly.

A central contribution of this work lies in its use of role-shifting as a design mechanism. By moving players from the position of the misunderstood subject to that of a mediating agent, the experience reframes empathy from passive sympathy to active responsibility. This transition may help explain why participants described the experience as meaningful within the study context, rather than merely emotionally moving. From an HCI perspective, these findings suggest that empathy-oriented systems may benefit from designs that foreground action and mediation, rather than observation alone.

The use of metaphor further enabled emotional engagement without didactic framing. By situating the experience in a non-human, marine context, the system created psychological distance that allowed participants to later map the experience onto real-world social situations, including immigration and marginalization. This supports prior HCI work arguing that metaphor and embodied interaction can facilitate reflective transfer across contexts, especially for sensitive social issues.

\textbf{Limitations.}
This study represents an early-stage evaluation with a limited sample size and short-term exposure. While the results indicate promising attitudinal shifts, the durability of these effects remains unknown. In addition, because the control condition involved documentary viewing without immersive interaction, the current design does not isolate the specific contribution of the role-shifting mechanism from broader effects of VR embodiment or experiential novelty. Future work could incorporate additional comparison conditions—such as a non-mediating VR vignette or a single-role immersive scenario—to more precisely examine the unique impact of role transition on empathic outcomes.

Future research should also examine long-term impact, compare alternative experience structures, and explore how such systems might integrate with real-world social or educational initiatives.

\section{Conclusion}
\textit{52-Hz Whale Song} illustrates how embodied, role-shifting VR experiences can support empathic engagement by transforming abstract social challenges into lived interactional experiences. By uniting metaphor, procedural interaction, and mediation-oriented design, this work contributes experience-driven design knowledge to HCI and motivates further exploration of interactive systems that foster empathy as action rather than observation.

\section{Acknowledgment of AI Use}
LLMs were used solely for minor language editing. 
All research content, analysis, and conclusions are entirely the authors’ own.

\bibliography{sample-base}

@article{wei2025systematic,
  title={Systematic literature review of using virtual reality as a social platform in HCI community},
  author={Wei, Xiaoying and Jin, Xiaofu and Lin Kan, Ge and Yan, Yukang and Fan, Mingming},
  journal={Proceedings of the ACM on Human-Computer Interaction},
  volume={9},
  number={2},
  pages={1--36},
  year={2025},
  publisher={ACM New York, NY, USA}
}

@article{bujic2025virtually,
  title={Virtually better: Multi-user experiment on avatar self-representation, self-discrepancies, avatar style and self-perceptions in a VR collaboration},
  author={Buji{\'c}, Mila and Macey, Anna-Leena and Kerous, Bojan and Buruk, O{\u{g}}uz and Hamari, Juho},
  journal={new media \& society},
  pages={14614448251323904},
  year={2025},
  publisher={SAGE Publications Sage UK: London, England}
}

@inproceedings{liang2025user,
  title={User Preferences for Interaction Timing in Smartwatch Sleep Hygiene Games},
  author={Liang, Zilu and Hwang, Daeun and Chen, Samantha and Hoang, Nhung Huyen and Khotchasing, Kingkarn and Melcer, Edward F},
  booktitle={Proceedings of the 2025 CHI Conference on Human Factors in Computing Systems},
  pages={1--17},
  year={2025}
}

@inproceedings{ozkaya2025investigating,
  title={Investigating the Motivational Game Elements in Game-based Interventions in School Context: A Literature Review},
  author={{\"O}zkaya, Mehmed Nihad and Baykal, G{\"o}k{\c{c}}e Elif},
  booktitle={Proceedings of the 2025 CHI Conference on Human Factors in Computing Systems},
  pages={1--16},
  year={2025}
}

@inproceedings{kirchner2024outplay,
  title={Outplay your weaker self: A mixed-methods study on gamification to overcome procrastination in academia},
  author={Kirchner-Krath, Jeanine and Schmidt-Kraepelin, Manuel and Sch{\"o}bel, Sofia and Ullrich, Mathias and Sunyaev, Ali and Von Korflesch, Harald FO},
  booktitle={Proceedings of the 2024 CHI Conference on Human Factors in Computing Systems},
  pages={1--19},
  year={2024}
}

@inproceedings{fernandez2025breaking,
  title={Breaking the Familiarity Bias: Employing Virtual Reality Environments to Enhance Team Formation and Inclusion},
  author={Fernandez-Espinosa, Mariana and Clouse, Kara and Sellars, Dylan and Tong, Danny and Bsales, Michael and Alcindor, Sophonie and Hubbard, Timothy D and Villano, Michael and G{\'o}mez-Zar{\'a}, Diego},
  booktitle={Proceedings of the 2025 CHI Conference on Human Factors in Computing Systems},
  pages={1--16},
  year={2025}
}

@inproceedings{wagener2025togetherreflect,
  title={TogetherReflect: Supporting Emotional Expression in Couples Through a Collaborative Virtual Reality Experience},
  author={Wagener, Nadine and Albensoeder, Daniel Christian and Reicherts, Leon and Wo{\'z}niak, Pawe{\l} W and Rogers, Yvonne and Niess, Jasmin},
  booktitle={Proceedings of the 2025 CHI Conference on Human Factors in Computing Systems},
  pages={1--16},
  year={2025}
}

@inproceedings{ju2025haptic,
  title={Haptic Empathy: Investigating Individual Differences in Affective Haptic Communications},
  author={Ju, Yulan and Meng, Xiaru and Taguchi, Harunobu and Gunasekaran, Tamil Selvan and Hoppe, Matthias and Ishikawa, Hironori and Tanaka, Yoshihiro and Pai, Yun Suen and Minamizawa, Kouta},
  booktitle={Proceedings of the 2025 CHI Conference on Human Factors in Computing Systems},
  pages={1--25},
  year={2025}
}

@article{gencc2024situating,
  title={Situating Empathy in HCI/CSCW: A Scoping Review},
  author={Gen{\c{c}}, U{\u{g}}ur and Verma, Himanshu},
  journal={Proceedings of the ACM on Human-Computer Interaction},
  volume={8},
  number={CSCW2},
  pages={1--37},
  year={2024},
  publisher={ACM New York, NY, USA}
}

@inproceedings{shrestha2025virtual,
  title={Virtual Worlds Beyond Sight: Designing and Evaluating an Audio-Haptic System for Non-Visual VR Exploration},
  author={Shrestha, Aayush and Malloch, Joseph},
  booktitle={Proceedings of the 2025 CHI Conference on Human Factors in Computing Systems},
  pages={1--19},
  year={2025}
}

@inproceedings{wang2025facilitating,
  title={Facilitating daily practice in intangible cultural heritage through virtual reality: A case study of traditional chinese flower arrangement},
  author={Wang, Yingna and Liu, Qingqin and Wei, Xiaoying and Fan, Mingming},
  booktitle={Proceedings of the 2025 CHI Conference on Human Factors in Computing Systems},
  pages={1--17},
  year={2025}
}

@inproceedings{desnoyers2025being,
  title={Being in Virtual Worlds: How Interaction Environment and Touch Shape Embodiment in Immersive Experiences},
  author={Desnoyers-Stewart, John and Antle, Alissa N and Riecke, Bernhard E},
  booktitle={Proceedings of the 2025 CHI Conference on Human Factors in Computing Systems},
  pages={1--16},
  year={2025}
}

@inproceedings{xie2025vrcaptions,
  title={VRCaptions: Design Captions for DHH Users in Multiplayer Communication in VR},
  author={Xie, Tianze and Zhang, Xuesong and Huang, Feiyu and Liu, Di and An, Pengcheng and Je, Seungwoo},
  booktitle={Proceedings of the 2025 CHI Conference on Human Factors in Computing Systems},
  pages={1--18},
  year={2025}
}

@inproceedings{kong2025working,
  title={Working Together Toward Interdependence: Chatbot-Based Support for Balanced Social Interactions Between Neurodivergent and Neurotypical Individuals},
  author={Kong, Ha-Kyung and Lowy, Rachel and Choi, Youjin and Kim, Jennifer G},
  booktitle={Proceedings of the 2025 CHI Conference on Human Factors in Computing Systems},
  pages={1--17},
  year={2025}
}

@inproceedings{choi2025aacesstalk,
  title={AACessTalk: Fostering Communication between Minimally Verbal Autistic Children and Parents with Contextual Guidance and Card Recommendation},
  author={Choi, Dasom and Park, SoHyun and Lee, Kyungah and Hong, Hwajung and Kim, Young-Ho},
  booktitle={Proceedings of the 2025 CHI Conference on Human Factors in Computing Systems},
  pages={1--25},
  year={2025}
}

@inproceedings{bei2024starrescue,
  title={StarRescue: the Design and Evaluation of A Turn-Taking Collaborative Game for Facilitating Autistic Children's Social Skills},
  author={Bei, Rongqi and Liu, Yajie and Wang, Yihe and Huang, Yuxuan and Li, Ming and Zhao, Yuhang and Tong, Xin},
  booktitle={Proceedings of the 2024 CHI Conference on Human Factors in Computing Systems},
  pages={1--19},
  year={2024}
}

@inproceedings{park2025lessons,
  title={Lessons from Real-World Settings: What Makes It Uniquely Difficult to Design Cognitive Training Programs for Children with Autism Spectrum Disorder and Other Developmental Disabilities},
  author={Park, Hyanghee and Jung, Sol Bee and Byun, Young Hee and Ahn, Daehwan and Park, Chan Woo and Byun, Sunjoo and Huang, Yun},
  booktitle={Proceedings of the 2025 CHI Conference on Human Factors in Computing Systems},
  pages={1--21},
  year={2025}
}

@inproceedings{smith2025archaeological,
  title={Archaeological Gameworld Affordances: A Grounded Theory of How Players Interpret Environmental Storytelling},
  author={Smith Nicholls, Florence and Cook, Michael},
  booktitle={Proceedings of the 2025 CHI Conference on Human Factors in Computing Systems},
  pages={1--20},
  year={2025}
}

@inproceedings{zhao2025immersive,
  title={Immersive Biography: Supporting Intercultural Empathy and Understanding for Displaced Cultural Objects in Virtual Reality},
  author={Zhao, Ke and Chen, Ruiqi and Zhang, Xiaziyu and Wang, Chenxi and Chen, Siling and Wang, Xiaoguang and Wang, Yujue and Tong, Xin},
  booktitle={Proceedings of the 2025 CHI Conference on Human Factors in Computing Systems},
  pages={1--17},
  year={2025}
}

@book{kolb2014experiential,
  title={Experiential learning: Experience as the source of learning and development},
  author={Kolb, David A},
  year={2014},
  publisher={FT press}
}

@book{bogost2010persuasive,
  title={Persuasive games: The expressive power of videogames},
  author={Bogost, Ian},
  year={2010},
  publisher={mit Press}
}

@inproceedings{chen2023design,
  title={Design and Evaluation of a VR Therapy for Patients with Mild Cognitive Impairment and Dementia: Perspectives from Patients and Stakeholders},
  author={Chen, Ruiqi and Wang, Shuhe and Xu, Xuhai and Wei, Lan and Sun, Yuling and Tong, Xin},
  booktitle={2023 IEEE Conference on Virtual Reality and 3D User Interfaces Abstracts and Workshops (VRW)},
  pages={597--598},
  year={2023},
  organization={IEEE}
}

@inproceedings{meng2026mistyforest,
  author    = {Yibo Meng and Bingyi Liu and Ruiqi Chen and Yan Guan},
  title     = {Misty Forest VR: Turning Real ADHD Attention Patterns into Shared Momentum for Youth Collaboration},
  booktitle = {Extended Abstracts of the 2026 CHI Conference on Human Factors in Computing Systems (CHI EA '26)},
  year      = {2026},
  publisher = {Association for Computing Machinery},
  address   = {New York, NY, USA},
  location  = {Barcelona, Spain},
  pages     = {1--7},
  doi       = {10.1145/3772363.3798689},
  isbn      = {979-8-4007-2281-3/2026/04}
}
\bibliographystyle{unsrt}
\end{document}